\documentstyle[twocolumn,aps,psfig,epsfig,floats]{revtex}
\begin{document}
\def\be{\begin{equation}}
\def\ee{\end{equation}}
\def\bc{\begin{center}}
\def\ec{\end{center}}
\def\bea{\begin{eqnarray}}
\def\eea{\end{eqnarray}}
\draft
\twocolumn[\hsize\textwidth\columnwidth\hsize\csname
@twocolumnfalse\endcsname
\title{Growing Random Networks with Fitness }
\author{G. Erg\"{u}n$^{*}$ and G. J. Rodgers}

\address{Department of Mathematical Sciences, Brunel University,
Uxbridge, Middlesex UB8 3PH, U.K.}
\address{$^{*}$Electronic address: Guler.Ergun@brunel.ac.uk}

\maketitle \thispagestyle{empty}
\begin{abstract}
   Three models of growing random networks with fitness dependent
   growth rates are analysed using the rate equations for the distribution
   of their connectivities. In the first model (A), a network
   is built by connecting incoming nodes to nodes of connectivity $k$
   and random additive fitness $\eta$, with rate $( k-1)+ \eta $.
   For $\eta >0$ we find the connectivity distribution is power law
   with exponent
   $\gamma=<\eta>+2$.  In the second model (B), the network is  built
   by connecting nodes to nodes of connectivity
   $k$, random additive fitness $\eta$ and random multiplicative fitness
   $\zeta$ with rate  $\zeta(k-1)+\eta$. This model also has a power
   law connectivity distribution, but with an exponent which depends
   on the multiplicative fitness at each node. In the third model (C),
   a directed graph is considered and is built by the addition of
   nodes and the creation
   of links.  A node with fitness $(\alpha, \beta)$, $i$ incoming
   links and $j$ outgoing
   links gains a new incoming link with rate $\alpha(i+1)$, and a new outgoing
   link with rate $\beta(j+1)$. The distributions of the number of
   incoming and outgoing links both scale as power laws, with  inverse
   logarithmic  corrections.
   \end{abstract}

]
\narrowtext

\section{Introduction}

Recently, there has been a considerable interest in the growth
properties of human interaction networks such as the world wide
web \cite{diameter,graphstructure}, the citation distribution of
publications \cite{citation}, the electrical distribution systems
\cite{barab} and the social networks \cite{bestconnected}. These
networks all have very different physical forms, with different
definitions for their nodes and links. However they appear to
display considerable topological similarity, having connectivity
distributions which behave as power laws. These distributions
cannot be explained by traditional random graph theory, which is
based on randomly connecting together a fixed number of nodes, and
results in Poisson distributions for the connectivity
\cite{boll,randomgraphs}.

Models of growing random graphs were first introduced by
Barab\'{a}si and Albert \cite{barab}, who identified two important
features that these graphs must possess in order to display power
law distributed connectivities. These features are  $(i)$ networks
grow by addition of new nodes and $(ii)$ new nodes preferentially
attach to highly connected nodes. Consideration of only these
elements in \cite{barab} led to the conclusion that large networks
can self-organize into a scale free state. Since then, many other
models
\cite{aging,exactsol,fitness,connectivity,heritable,amaral,organization,scalefree,scaling,condensation}
have emerged to study various properties of these graphs such as
aging \cite{aging,amaral}, connectivity \cite{connectivity},
inheritance \cite{heritable}, permanent deletion of links and
nodes \cite{scaling} and their effect on a growing network
topology. The main conclusion of all these models is that
incorporation of additional features changes the scaling behaviour
of growing random networks. However it is still not understood why
most of the empirical work observes power law exponents between 2
and 3, and the analytical work recovers exponents that range
between 2 and $\infty$ \cite{aging,organization}. Furthermore,
some of the more detailed features of the networks have not yet
been captured \cite{graphstructure,rodgers}.

In this paper, based on an idea introduced by \cite{fitness} we
study the influence of quenched disorder which we call {\it
fitness}, on the growth rates of networks. Similar ideas have been
studied in other models, either through  the initial
attractiveness of a node \cite{exactsol} or the fitness of a site
to compete for links \cite{fitness}. However, our approach is
somewhat different to these models.  We use a rate equation
approach \cite{connectivity} to generalize and solve three network
models with different growth rates.

In Sec. \ref{sec:model A}, we investigate the effect of additive
randomness, while the effect of multiplicative randomness is
analyzed in Sec. \ref{sec:model B}. In Sec. \ref{sec:model C}, we
assume the network is a directed graph
\cite{randomgraphs,dgraph,rodgers,ken} and both incoming and
outgoing links are considered, to model the growth of the world
wide web. We summarize our results and draw conclusions in the
last section.
\section{Model A}
\label{sec:model A}

In this model, we consider a network where a fitness $\eta$,
chosen from a probability distribution $f_A (\eta)$, is assigned
to each node. The network is built by connecting incoming nodes to
nodes of connectivity $k$ and fitness $\eta$ with rate $(k-1)+
\eta$, that is to say, there is a linear preferential attachment
to nodes with already high number of links and a high fitness
$\eta$. This simply means that not all nodes which $k$ existing
links are equivalent because $k$ does not enclose the full
information about the popularity of a node. For instance, if a
node is a web site, $\eta$ could be a measure of the number of
related TV commercials, or tube advertisements.  Using the rate
equation approach we describe the time evolution of the average
number of nodes of connectivity $k$ and fitness $\eta$,
$N_{k}(\eta)$, as

\begin{eqnarray}
\label{Ank} \frac{\partial N_{k}(\eta)}{\partial t}&=&\frac{1}{M}
\left[(k+\eta-2)N_{k-1}(\eta)-(k+\eta-1)N_{k}(\eta)\right]\nonumber
\\ &&+ \delta_{k1}f_A (\eta).
\end{eqnarray}
The first term on the right hand side of Eq. (\ref{Ank})
represents the increase in the number of sites with $k$ links when a site with
$k-1$ links gains a link. The second term expresses the loss of
sites with $k$ links when they gain a new link. The last term
accounts for the continuous addition of nodes of connectivity $1$
and fitness $\eta$ with probability $f_A (\eta)$. The
multiplicative factor $M$ is defined by

\begin{equation} \label{ANorm}
M(t)=\sum_{k, \eta}(k+ \eta -1) N_{k}(\eta),
\end{equation}
which ensures that the equation is properly normalized. Before
going any further, let us make some remarks. First, to obtain a
growing network, we need $k-1+\eta >0$ for all $k$, so that $\eta
>0$, because from the definition of the model, each site is
created with one link. Second, all sites associated with $\eta =
1$ have the simple linear preferential attachment of earlier
models \cite{barab,connectivity,organization}. Finally, $f_A (\eta
)$ can either be discrete or continuous.

We analyse the model from the rate equation starting with the
moments of $N_{k}(\eta)$ defined by

\begin{eqnarray}
\label{AMoments}
M_{ij}(t)&\equiv&\sum_{k,\eta}k^{i}\eta^{j}N_k(\eta).
\end{eqnarray}
We can easily show that

\begin{eqnarray}
\frac{\partial M_{00}}{\partial t}&=&1, \quad \frac{\partial
M_{10}}{\partial t}=2\quad \hbox{and} \quad \frac{\partial
M_{01}}{\partial t}=<\eta>
\end{eqnarray}
where $<\eta>$ is the average value of the fitness.
For large times, the initial values of the moments become
irrelevant, so that we get

\begin{eqnarray}
M(t)&=&M_{10}(t) + M_{01}(t)- M_{00}(t)=[<\eta>+1]t.
\end{eqnarray}
Similarly, it can be shown that $N_{k}(\eta ,t)$ and all its
moments grow linearly with time. Therefore, we can write
$N_{k}(\eta,t)=tn_{k}(\eta)$ and $M(t)=mt$. The latter relation
implies $m = <\eta>+1$, while we insert the former in
Eq.(\ref{Ank}) to obtain the recurrence relation

\begin{equation}
\label{Arecursion} (k+\eta +m -1) n_k(\eta)=(k+\eta
-2)n_{k-1}(\eta)+m\delta_{k1}f_A (\eta).
\end{equation}
Solving Eq.~(\ref{Arecursion}), we obtain

\begin{eqnarray}
n_{k}(\eta)&=&\frac{\Gamma(k+ \eta -1)}{\Gamma (k+ \eta +m)}
\frac{\Gamma(\eta +m)}{\Gamma (\eta)} m f_A (\eta).
\label{eq:nkmodel a}
\end{eqnarray}
In particular, the rate of change of the total number of links
connected to the sites with fitness $\eta$ is equal to

\begin{eqnarray}
\sum_{k=1}^{\infty}
kn_{k}(\eta)&=&\left(1+\frac{\eta}{<\eta>}\right)f_A (\eta).
\end{eqnarray}
For large $k$, Eq. (\ref{eq:nkmodel a}) is equivalent to

\begin{equation}
\label{asymptotic} n_{k} (\eta)\sim k^{-(m+1)}\, \sim
k^{-(<\eta>+2)}.
\end{equation}
The distribution scales as a power law $n_{k}(\eta)\sim
k^{-\gamma}$ with an exponent $\gamma= <\eta>+2$, which depends
only on the average fitness $<\eta>$, and consequently is the same
for every node. Hence, the introduction of an additive random
fitness at each node, modifying the preferential attachment
process, generates a power law connectivity distribution. The
exponent of this power law is shifted by $<\eta> -1$ with respect
to its value when the preferential attachment is simply linear. Of
course, for $f_A (\eta)= \delta(\eta-1)$,
\begin{equation}
n_{k}\sim k^{-3},
\end{equation}
which is, as expected, the result obtained in  previous models
without fitness \cite{barab,connectivity,organization}.

\section{Model B}
\label{sec:model B}

In the previous section, we introduced a model where linear
preferential attachment is decorated by a random additive process
to construct an independent source of preferential attachment.
However, even if it seems reasonable to assume that the attachment
is proportional to the number of already existing links, there is
no specific reason to assume that the coefficient of
proportionally is the same for every node. In this section, we
consider a network where each node is associated to a triplet $(k,
\eta,\zeta)$. The network is built by adding a new node at each
time step and connecting it to a node with random additive fitness
$\eta$, random multiplicative fitness $\zeta$ and connectivity $k$
with rate $\zeta(k-1)+\eta$. Where $\eta$ and $\zeta$ are quenched
variables, initially chosen from a probability distribution $f_B
(\eta, \zeta)$. This model is a generalisation of that introduced
in \cite{fitness}, and Model A is
recovered when $f_B (\eta, \zeta) = f_A (\eta)\delta (\zeta -
1)$. The multiplicative fitness symbolizes the fact that, even if
the growth rate is proportional to already existing links, there
can exist different categories of nodes which attract new links at
different rates.

The rate equation for this model, which describes the time
evolution of the average number of nodes with triplet $(k,
\eta,\zeta)$, $N_{k}(\eta, \zeta)$, is given by

\begin{eqnarray}
\label{Bnk} \frac{\partial N_{k} (\eta,\zeta)}{\partial t} &=&
\frac{1}{M} \left([\zeta(k-2)+\eta ]N_{k-1}(\eta,
\zeta)\right.\nonumber
\\ && \left.-[\zeta(k-1)+ \eta ]N_{k}(\eta,\zeta)\right)
\,+\delta_{k1}f_B (\eta,\zeta ).
\end{eqnarray}
The terms on the right-hand side of this equation are analogous to
those in Eq.~(\ref{Ank}), with the new preferential rates of
growth. The normalization factor here is

\begin{equation}
M(t)=\sum_{k,\eta,\zeta}\left[\zeta(k-1)+\eta\right]N_{k}
(\eta,\zeta). \label{eq:bigM}
\end{equation}
To solve Eq.~(\ref{Bnk}), we employ the same technique as in the
previous section, defining the moments of $N_{k}(\eta,\zeta)$ by

\begin{equation}
M_{ijl} \equiv \sum_{k,\eta,\zeta} k^i \eta^j \zeta^l
N_{k}(\eta,\zeta).
\end{equation}
Looking at the lowest moments of $N_{k}(\eta,\zeta)$, we find

\begin{equation}
\label{Bmoments}
  \frac{\partial M(t)}{\partial t} =
\frac{1}{M}\sum_{k,\eta,\zeta}\zeta [\zeta(k-1)+\eta]
  N_{k}(\eta,\zeta)+<\eta>,
\end{equation}
where  $<\eta>$ is the average additive fitness. Again, it is easy
to prove that $N_{k}(\eta,\zeta,t)$ and all its moments are linear
functions of time. Hence, we define $m$ and $n_{k}(\eta,\zeta)$
through $M(t)\equiv mt$ and $N_{k}(\eta,\zeta,t)\equiv
tn_{k}(\eta,\zeta)$, respectively. We refer to $m$ as the reduced
moment from now on.

Eq.~(\ref{Bmoments}) implies that the reduced moment is a solution
of
\begin{equation}
\label{littlem} m=\frac{1}{m}\sum_{k, \eta,
\zeta}\zeta[\zeta(k-1)+\eta]n_{k}(\eta,\zeta)+<\eta>.
\end{equation}
 From Eq.~(\ref{Bnk}), we obtain

\begin{eqnarray}
\label{Brecursion}
[\zeta(k-1)+\eta+m]n_{k}(\zeta,\eta)&=&[\zeta(k-2)+\eta]n_{k-1}(\zeta,\eta)
\nonumber \\ &&+ m\delta_{k1}f_B(\zeta,\eta ).
\end{eqnarray}
The previous relation yields

\begin{equation}
\label{Bgamma}
n_{k}(\eta,\zeta)=\frac{\Gamma\left(k+\frac{\eta}{\zeta}-1\right)}
            {\Gamma\left(k+\frac{\eta +m}{\zeta}\right)}
            \frac{\Gamma\left(\frac{\eta +m}{\zeta}\right)}
            {\Gamma\left(\frac{\eta }{\zeta}\right)}
            \frac{m}{\zeta}f_B(\eta,\zeta ).
\end{equation}
When $k \, \to \infty $,
\begin{equation}
n_{k}(\eta,\zeta)\sim k^{-\left(1+\frac{m}{\zeta} \right)}.
\end{equation}
The growth rate of the number of sites associated with a triplet
$(k, \eta,\zeta)$, scales asymptotically as a power law $n_k \sim
k^{-\gamma}$, with an exponent $ \gamma = 1+m/\zeta$, which
depends on the fitness at a particular site. It means that, unlike
the additive fitness, the multiplicative fitness generates
multiscaling, with a different power law for each fitness.

To complete the solution of the model, we need to obtain an
expression for the reduced moment, $m$. For this purpose, we
introduce a generating function defined as

\begin{eqnarray}
g(x,\eta,\zeta)&\equiv&\sum^\infty_{k=1}x^k n_{k}(\eta,\zeta ).
\end{eqnarray}
Eq. (\ref{Brecursion}) gives

\begin{eqnarray}
g(1,\eta,\zeta)&=&f_B (\eta,\zeta )=\sum_{k=1}^{\infty}
n_{k}(\eta,\zeta )
\end{eqnarray}
and
\begin{eqnarray}
g'(1,\eta,\zeta)&=&\frac{\eta -\zeta+m}{m-\zeta}f_B(\eta,\zeta
)=\sum_{k=1}^{\infty} k n_{k}(\eta,\zeta).
\end{eqnarray}
Substituting in Eq.~(\ref{littlem}) leads to an implicit equation
for $m$,

\begin{equation}
\label{unity} \int f_B (\eta,\zeta )\frac{\eta }{m-\zeta}\,d
\eta \,d\zeta = 1
\end{equation}
which cannot be solved explicitly.
We can define $n_k$, the
connectivity distribution of the entire network, as

\begin{equation}
n_k \equiv \int f_{C}(\eta, \zeta) n_k (\eta, \zeta) d\eta d\zeta.
\end{equation}

As an example, we consider $f_B (\eta,\zeta )=1$, $0\leq \eta\leq
1$ and $0\leq \zeta \leq 1$. Solving Eq.~(\ref{unity}) gives $m=
1/(1-e^{-2}) = 1.156$ and, integrating Eq.~(\ref{Bgamma}) over
$\eta$ and $\zeta$ within the chosen limits,

\begin{eqnarray} \label{doubleint} n_{k} &=&
\int^1_{0}\!\!\!\int^1_{0}k^{-\left(\frac{m}{\zeta}+1\right)}
\frac{\Gamma\left(\frac{\eta +m}{\zeta}\right)}
{\Gamma\left(\frac{\eta }{\zeta}\right)}\frac{m}{\zeta}\,d\eta
d\zeta.
\end{eqnarray}
We find that the connectivity distribution in the asymptotic limit
$k \to \infty$ is

\begin{equation}
\label{Bdistribution}
 n_{k}\sim \frac{1}{\ln k}k^{-(1+m)}.
\end{equation}
This is simply a power law form multiplied with an inverse
logarithmic correction and substitution of $m$ yields

\begin{equation} n_{k}\sim \frac{1}{\ln k}k^{-2.156}.
\end{equation}
By using $f_B (\eta,\zeta )=\delta(\zeta-\eta)$ with $0\leq
\zeta\leq1$, the solution obtained by \cite{fitness} is recovered,
which has the same functional form as Eq.~(\ref{Bdistribution})
with a power law exponent $\gamma=2.255$.

\begin{figure}
\centerline{\epsfig{file=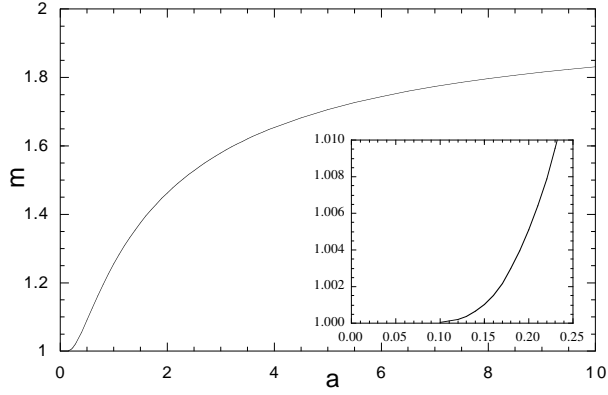,width=8.5cm}} \caption{$m$
against $a$ for Model B when the multiplicative fitness is
distributed as a power law with exponent $a-1$. The inset is the region close to the origin.} \label{fig1}
\end{figure}

We can solve for the large $k$ behaviour of the connectivity distribution for
a number of different forms of the fitness. For instance, if

\begin{equation}
f_B (\eta,\zeta )= a\zeta^{a-1}\delta(\zeta-\eta)
\end{equation}
and $0\leq\zeta\leq1$, $a>0$, then the connectivity distribution
behaves as

\begin{equation}
n_{k}\sim\frac{1}{\ln k}k^{-(m+1)}
\end{equation}
as $k\to \infty$ and $m(a)$ satisfies

\begin{equation}
\label{ma} \int^1_{0}\frac{a\zeta^a}{m-\zeta}d\zeta =1.
\end{equation}
Using this equation it is simple to show that as $a\to0$, $m\to1$ and as
$a\to\infty$, $m\to2$. In Fig.~(\ref{fig1}) this equation is
solved numerically.
We find that $1\leq m\leq
2$, implying that the power law exponent is in the range of (2,3) which is in
very good agreement with the experimental results
\cite{barab,citation,graphstructure}.

Another example to consider is when

\begin{equation}
f_{B}(\eta,\zeta)=6\zeta(1-\zeta)\delta(\zeta-\eta)
\end{equation}
with $0\leq \zeta \leq 1$. In this case the connectivity distribution takes
the form

\begin{equation}
n_{k}\sim\frac{1}{(\ln k)^{2}}k^{-(m+1)}.
\end{equation}
as $k\to \infty$ where $m=1.550$.

\section{Model C}
\label{sec:model C}

In the previous two models, the links were undirected and the
number of links and nodes were equal, which is not a good model
for some growing networks such as the www. In this section, a
directed network is built by node and link addition. At each time
step, with probability $p$, a new node is added and with
probability $q=1-p$, a new directed link is created between two
nodes. A node with fitness $(\alpha,\beta)$, $i$ incoming links
and $j$ outgoing links, will gain a new incoming link with rate
$\alpha(i+1)$ and a new outgoing link with rate $\beta(j+1)$.
Then, the connectivity distribution $N_{ij}(\alpha,\beta)$, the
average number of nodes with $i$ incoming and $j$ outgoing links,
evolves as

\begin{eqnarray}
\label{nij} \frac{\partial N_{ij}}{\partial t}(\alpha,\beta)
&=&\frac{q\alpha}{M_{1}}\left[iN_{i-1j}(\alpha,\beta)-(i+1)N_{ij}
(\alpha,\beta)\right]\nonumber \\
&&+\frac{q\beta}{M_{2}}\left[jN_{ij-1}(\alpha,\beta)-(j+1)N_{ij}
(\alpha,\beta)\right] \nonumber \\
&&+p\,\delta_{i0}\delta_{j0}\,f_{C}(\alpha ,\beta ).
\end{eqnarray}
The first term in the first square brackets represents the
increase of $N_{ij}$ nodes when nodes with $i-1$ incoming and $j$
outgoing links, gain an incoming link and the second term
represents the corresponding loss. The second square brackets
contain the analogous terms for outgoing links and the last term
ensures the continuous addition of new nodes with fitness $\alpha,
\, \beta$ with probability $f_{C}(\alpha,\beta)$. $M_{1}$ and
$M_{2}$ are the normalization factors, given by

\begin{eqnarray}M_{1}&=&\sum_{ij \alpha
\beta }(i+1)\alpha N_{ij}(\alpha,\beta)\quad \text{and}\\
M_{2}&=&\sum_{ij \alpha \beta }(j+1)\beta N_{ij}(\alpha,\beta).
\end{eqnarray}

From the definition of the model, one has
\begin{equation}
\sum_{ij \alpha \beta}N_{ij} (\alpha,\beta)=pt,
\end{equation}
which simply states that nodes are added with probability $p$. We
also have

\begin{eqnarray}
\sum_{ij \alpha \beta}\alpha N_{ij} (\alpha,\beta)&=& p<\alpha>t
\\ \sum_{ij \alpha \beta}\beta N_{ij} (\alpha,\beta)&=& p<\beta>t.
\end{eqnarray}
We define $n_{ij}(\alpha,\beta)$ through
$N_{ij}(\alpha,\beta,t)\equiv t n_{ij}$, where from now on we drop
the explicit $(\alpha,\beta)$ dependence to ease the notation.
Also, we can define the reduced moments $m_1$ and $m_2$ by
$M_{1}(t) \equiv tm_{1}$ and $M_{2}(t) \equiv tm_{2}$. Hence we
have

\begin{eqnarray} \label{cm1}m_{1}=\frac{q}{m_{1}}\sum_{ij \alpha
\beta }\alpha ^2(i+1)n_{ij}+p<\alpha > &&
\end{eqnarray} and similarly

\begin{eqnarray}\label{cm2} m_{2}=\frac{q}{m_{2}}\sum_{ij\alpha
\beta }\beta ^2(j+1)n_{ij}+p<\beta >.&&
\end{eqnarray}
 From Eq.~(\ref{nij}), we obtain

\begin{eqnarray}
\label{recursionc}
   &&n_{ij}[m_{1}m_{2}+m_{1}q\beta(j+1)+m_{2}q\alpha(i+1)]\nonumber
\\ &&= m_{2}\,q \,\alpha\,in_{i-1j} +m_{1}\,q\,\beta \,j n_{ij-1}\nonumber
\\ &&+ m_{1}\,m_{2}\,p\,\delta_{i0}\delta_{j0}\,f_{C}(\alpha ,\beta ).
\end{eqnarray}

Now, we consider the incoming link distribution
\begin{eqnarray}g_{i}=\sum_{j=0}^{\infty} n_{ij}
\end{eqnarray}
and the outgoing link distribution
\begin{eqnarray} h_{j}=\sum_{i=0}^{\infty} n_{ij}.&&
\end{eqnarray}
 From Eq. (\ref{recursionc}), the recurrence relations

\begin{equation}
\left( i + 1 + \frac{m_1}{q\alpha }\right) g_i = i g_{i-1} +
\frac{m_1 p}{q\alpha} \delta_{i0} f_{C}(\alpha,\beta)
\end{equation}
and

\begin{equation}
\left( j + 1 + \frac{m_2}{q\beta }\right) h_j = j h_{j-1} +
\frac{m_2 p}{q\beta} \delta_{j0} f_{C}(\alpha,\beta)
\end{equation}
are obtained. Solving these gives the incoming and the outgoing
links distributions

\begin{equation}
\label{incoming} g_{i} =
\frac{\Gamma(i+1)\Gamma(\frac{m_{1}}{q\alpha}+1)}{\Gamma(\frac{m_{1}}{q\alpha}+i+2)}
   \frac{m_{1}p}{q\alpha} f_{C}(\alpha,\beta)
\end{equation}
and

\begin{equation}
\label{outgoing} h_{j} =
\frac{\Gamma(j+1)\Gamma(\frac{m_{2}}{q\beta}+1)}{\Gamma(\frac{m_{2}}{q\beta}+j+2)}
   \frac{m_{2}p}{q\beta} f_{C}(\alpha,\beta).
\end{equation}
In the asymptotic limit both distributions are power laws; as $i
\to \infty$, $g_i \sim i^{-\gamma_{in}}$ with
$\gamma_{in}=(1+m_{1}/q\alpha)$ and for $j \to \infty $, $h_j \sim
j^{-\gamma_{out}}$ with $\gamma_{out}=(1+m_{2}/q\beta)$.

The appearance of both multiplicative fitnesses $\alpha$ and
$\beta$ in the exponents of the power laws, reflects the fact that
growing networks such as the www are evolving on the basis of
competition. The fitnesses here, can be thought of as a measure of
attractiveness of the content of a web page. This means that
within a particular commercial sector on the web, such as search
engines, e-mail account providers, the software design, films,
music and specific information,  the fittest competitors have
managed to gather millions of registered users in a very short
span of time.

To express these exponents numerically we will find implicit
equations for $m_{1}$ and $m_{2}$. Therefore, we use a generating
function defined as

\begin{equation}
g(x,y, \alpha,\beta )=\sum_{i,j=0}^{\infty} x^i y^j n_{ij}
\end{equation}
to find equations for $m_1$ and $m_2$. We have

\begin{equation}
g(1,1)=pf_{C}(\alpha,\beta)
\end{equation}
and

\begin{equation}
\left. \frac{\partial g}{\partial x} \right|_{x=y=1}
=\frac{pq\alpha}{m_{1}-q\alpha}f_{C}(\alpha,\beta).
\end{equation}
Hence

\begin{equation}
\sum_{i,j =0}^{\infty} in_{ij} = \frac{pq\alpha }{m_{1}-q\alpha}
f_{C}(\alpha ,\beta )
\end{equation}
and by an identical method

\begin{equation}
\sum_{i,j =0}^{\infty} jn_{ij} = \frac{pq\beta }{m_{2}-q\beta
}f_{C}(\alpha,\beta).
\end{equation}
Substitution of the above relations into Eq.~(\ref{cm1}) and
Eq.~(\ref{cm2}) gives implicit equations for $m_1$ and $m_2$,

\begin{equation}
p \sum_{\alpha, \beta} \frac{\alpha f_{C}(\alpha,\beta
)}{m_{1}-q\alpha } = 1 \label{eq:m1}
\end{equation}
and

\begin{equation}
p \sum_{\alpha, \beta} \frac{\beta f_{C}(\alpha,\beta
)}{m_{2}-q\beta } = 1, \label{eq:m2}
\end{equation}
respectively. The summations run over all possible values of
$\alpha$ and $\beta$ and can be replaced by integrations for
continuous distributions.

First, we consider a general case; if
\begin{equation}
f_{C}(\alpha,\beta)=f_{C}(\beta,\alpha)
\end{equation}
then the distribution of incoming and outgoing links is the same,
$g_{i}=h_{i}$.

As with Model B, we will consider two particular non-trivial
distributions of the fitness. For power law fitnesses

\begin{equation}
f_{C}(\alpha,\beta)=ab\alpha^{a-1}\beta^{b-1}
\end{equation}
we find that the distribution of incoming links is given by

\begin{equation}
\label{gi} g_{i}\sim\frac{1}{\ln i}i^{-(1+\frac{m_1}{q})}
\end{equation}
and analogously for outgoing links

\begin{equation}
h_{j}\sim\frac{1}{\ln j}j^{-(1+\frac{m_2}{q})}
\end{equation}
for large $i$ and $j$. The parameter $m_1$ is a function of both
$a$ and $p$ and $m_2$ is a function of $b$ and $p$. It is a simple
matter to show that $m_1(a)=g(a)$ and $m_2(b)=g(b)$ where $g(c)$
satisfies

\begin{equation}
\label{g} p \int^1_{0}\frac{c\,x^c}{g(c)-qx}dx =1.
\end{equation}
Consequently, as $a \to 0$, $m_1 \to q$ and as $a\to\infty$,
$m_1\to 1$. Thus by picking $a$ appropriately, the power law in
the distribution of incoming links Eq.~(\ref{gi}) can have an
exponent with any value between 2 and $1+1/q$. A similar situation
occurs with $m_2(b)$.

The fitnesses for incoming and outgoing links can be more strongly
coupled together. An example of this is

\begin{equation}
\label{fc} f_{C}(\alpha,\beta) = \cases{
                  2 & $\alpha > \beta$,\cr
                  0 & $\beta > \alpha$}
\end{equation}
where we find that for large $i$ and $j$, $g_{i}$ has the same
form as Eq.~(\ref{gi}) and

\begin{equation}
h_{j}\sim \frac{1}{(\ln j)^{2}}j^{-(1+\frac{m_{2}}{q})}.
\end{equation}
The stronger coupling between the fitnesses is reflected in the
different functional forms of the probability distributions for
incoming and outgoing links.

\section{Discussion and conclusions}
\label{sec:conclusions}

We have studied three growing network models with the
consideration of two key elements in mind;  (i) networks are
continuously growing, (ii) the attachment process is preferential.
In the first model in Sec. \ref{sec:model A}, we found that the
introduction of random additive fitness (quenched disorder) $\eta$
at each node modified the preferential attachment process and the
generated network had a power law connectivity distribution with
an exponent $\gamma=<\eta>+1$, where only the average value of the
fitness is of importance. However, introduction of both random
additive fitness $\eta$ and random multiplicative fitness $\zeta$
in Sec. \ref{sec:model B}, led to a scale free network where the
exponent $\gamma=1+m/\zeta$ depends on the fitness $\zeta$ at each
node. When the fitnesses were distributed with a power law
distribution between 0 and 1, the connectivity distribution of the
whole system was power law  with a logarithmic correction, with
the value of the exponent in the power law between 2 and 3.

In Sec. \ref{sec:model C} we studied a directed graph which was
allowed to form  loops in an attempt to model a different class of
growing random graphs. The incoming and the outgoing link
distributions exhibit power law forms with exponents corresponding
to $\gamma_{in}=(1+m_{1}/q\alpha)$ and
$\gamma_{out}=(1+m_{2}/q\beta)$ depending upon the values of the
fitness at each site. Choosing a particular fitness distribution
and calculating the connectivity distribution for the whole system
often results in power laws mediated by logarithmic corrections.
We gave two examples of such behaviour.

There are a great many examples of random growing graphs in
science, social science, technology and biology. Only a fraction
of these systems have been characterised experimentally. Whilst
the systems studied in this paper are not as theoretically
appealing as those with pure power law forms, it seems likely that
some of the real random growing networks will be described by
models of this particular type.

\acknowledgments We would like to thank the EPSRC and The
Leverhulme Trust for financial support and Ren\'e D'Hulst for
useful discussions.


\end{document}